\documentclass[a4paper,epj]{svjour}

\usepackage{graphicx}
\usepackage{amsmath}
\usepackage{amssymb}
\usepackage{cite}
\usepackage{units}

\begin{document}

\title{Sidebands in the light absorption of driven metallic nanoparticles}
\author{Guillaume Weick\inst{1,2,3} \and Gert-Ludwig Ingold\inst{2} 
\and Dietmar Weinmann\inst{1} \and Rodolfo A.\ Jalabert\inst{1,2}}
\institute{
  Institut de Physique et Chimie des Mat\'eriaux de Strasbourg, 
  UMR 7504 (ULP-CNRS), 23 rue du Loess, BP 43, F-67034 Strasbourg Cedex 2,
  France
  \and 
  Institut f\"ur Physik,
  Universit\"at Augsburg, Universit\"atsstra{\ss}e 1, D-86135
  Augsburg, Germany
  \and 
  Fachbereich Physik, Freie Universit\"at Berlin, Arnimallee 14, 
  D-14195 Berlin, Germany
}

\date{\today}

\abstract{The dynamics of the surface plasmon in laser-driven metallic 
nanoparticles is described by means of a master-equation formalism. Within
the Markov approximation, the dynamics is studied for different regimes
ranging from weak excitation in photoabsorption experiments to strong 
excitation in pump-probe spectroscopy. It is shown that two collective levels
are sufficient to describe the dynamics of the surface plasmon. On this basis,
we predict the appearance of sidebands in the absorption spectrum of the
probe laser field in pump-probe experiments.
  \PACS{%
    {78.67.Bf}{Nanocrystals and nanoparticles} \and
    {73.20.Mf}{Collective excitations} \and
    {71.45.Gm}{Exchange, correlation, dielectric and magnetic response
    functions, plasmons} 
  }
}

\maketitle

%==============================================================================
%==============================================================================
\section{Introduction}
A surface plasmon excited by a pump laser in a metallic nanoparticle 
decays on the scale of tens of femtoseconds by heating the electron
system \cite{deheer,brack_RMP}. The equilibration with the ionic lattice 
occurs on the much longer timescale of a few picoseconds. This electronic 
relaxation has been thoroughly studied by means of pump-probe experiments 
\cite{tokiz94, bigot1, perner, halte99}. Present femtosecond resolution 
experiments allow to concentrate on the initial dynamics of the electronic 
degrees of freedom \cite{bigot2, delfatti, voisin}. While the dynamics on 
the picosecond scale has been analysed with quasi-equilibrium theories 
assuming the thermalisation of the electron gas as a whole \cite{lett}, 
much less is known for the initial dynamics. It is precisely this initial 
regime on which we focus in this work.

For the description of the initial dynamics of the electron gas, 
before it has thermalised after an excitation by a laser pulse, 
it is appropriate to decompose the electron system into two parts
\cite{gerchikov,weick05,weick06}. 
The main degree of freedom describes the surface plasmon and refers to
the centre of mass of the electron gas. The remaining degrees of freedom
consist of the relative coordinates. If their number is sufficiently
large, this electronic subsystem behaves to a good approximation as a heat 
bath to which one can attribute an effective temperature.

The surface plasmon coupled to the relative coordinates and driven by 
an external field is reminiscent of the situation encountered in quantum 
optics, where a laser-driven atom is coupled to an environment consisting 
of the electromagnetic field modes \cite{cohen}. This analogy leads us
to predict the appearance of sidebands in the light absorption of a weak 
probe beam while the nanoparticle is illuminated by a strong pump laser. 
This effect should be experimentally observable and offers a 
possibility to study the surface plasmon dressed by photons of the
pump laser field. 

We start in the next section by introducing the model of a laser-driven 
nanoparticle. In Section~\ref{sec_RDM}, a 
master-equation description for the dynamics of the electronic centre of 
mass coupled to the relative-coordinate system is developed. We will argue 
that the Markov approximation applies in our situation. In order to make
progress, we restrict in Section~\ref{sec_2l} our description of the
surface plasmon to its two lowest eigenstates and discuss the resulting
dynamics. The regime of validity of the two-level approximation will be 
discussed in the appendix. In Section~\ref{sec_s}, the model parameters for 
typical experimental situations are estimated. Finally, we present in 
Section~\ref{sec_abs} our prediction of sidebands in the absorption spectrum 
of a weak probe beam impinging on a nanoparticle subject to a pump laser.

%==============================================================================
%==============================================================================
%==============================================================================
%==============================================================================
\section{Model of a laser-driven nanoparticle}
\label{sec_model}
The electron gas formed by the $N$ valence electrons of the metallic 
nanoparticle is described by the Hamiltonian
\begin{equation}
\label{H}
H = \sum_{i = 1}^{N} \left[ \frac{p_i^2}{2 m_{\rm e}} + U(|{\bf r}_i|) \right]
+ \frac{e^2}{2} \sum_{\substack{i, j = 1\\(i \neq j)}}^{N}
\frac{1}{\left| {\bf r}_i - {\bf r}_j \right|}\,,
\end{equation}
where ${\bf r}_i$ is the position of the $i^{\rm th}$ electron. The
single-particle confining potential $U$ represents the interaction of the 
electrons with the ionic background which within the jellium model is 
approximated by a uniformly charged sphere of radius $a$ \cite{brack_RMP}. 
As a function of the radial coordinate, this central potential is harmonic 
for $|{\bf r}|<a$ and Coulomb-like for $|{\bf r}|>a$.

For our purposes it is useful to express the Hamiltonian \eqref{H} in terms 
of the centre-of-mass and relative coordinates as \cite{gerchikov,weick06} 
\begin{equation}
\label{H_decomposition}
H = H_{\rm cm}+H_{\rm rel}+H_{\rm c}\,.
\end{equation}
For small centre-of-mass displacements, $H_{\rm cm}$
is well described in terms of the usual ladder operators  
$b^\dagger$ and $b$ by the harmonic oscillator Hamiltonian
\begin{equation}
\label{H_cm}
H_{\rm cm} = \hbar \tilde\omega_{\rm M} b^\dagger b 
\end{equation}
with a frequency
\begin{equation}
\label{omega_M}
\tilde\omega_{\rm M}=\omega_{\rm M}\sqrt{1-\frac{N_{\rm out}}{N}}\,.
\end{equation}
Here, $\omega_{\rm M} = (4\pi n_{\rm e} e^2/3 m_{\rm e})^{1/2}$ is the
classical Mie fre\-quen\-cy  which depends on the charge $e$, the mass $m_{\rm e}$ 
and the bulk density $n_{\rm e}$ of the electrons. The spill-out effect 
\cite{brack_RMP}, which accounts for the fact that a fraction $N_{\rm out}/N$ 
is found outside the nanoparticle, reduces the electron density and thus 
leads to a redshift of the plasmon frequency from the Mie frequency to 
$\tilde\omega_{\rm M}$. Except for extremely small clusters, anharmonicities
yield negligible corrections to \eqref{H_cm} and are thus disregarded here
\cite{hagino}.

Within the mean-field approximation, the Hamiltonian for the relative
coordinates reads
\begin{equation}
\label{H_rel}
H_{\rm rel}=\sum_\alpha\varepsilon_\alpha c_\alpha^\dagger c_\alpha\,, 
\end{equation}
where $\varepsilon_\alpha$ are the eigenenergies in the effective mean-field 
potential $V$ and $c_\alpha^{\dagger}$ ($c_\alpha$) are the creation 
(annihilation) operators associated with the corresponding one-body 
eigenstates $|\alpha\rangle$. Numerical calculations within the local 
density approximation show that, for the purpose of analytical calculations,
$V$ can in general be approximated by a step-like potential at the surface
of the nanoparticle \cite{weick05,weick06}.

The linearised coupling Hamiltonian between the cen\-tre-of-mass and
relative-coordinate systems can be written as  
\begin{equation}
\label{H_c}
H_{\rm c} =  \Lambda\left(b^\dagger + b\right)\sum_{\alpha\beta}d_{\alpha\beta}
c_\alpha^{\dagger}c_\beta\,,
\end{equation}
where the coupling strength $\Lambda=(\hbar m_{\rm e}\omega_{\rm
M}^3/2N)^{1/2}$. 
The matrix element $d_{\alpha\beta}$ between two eigenstates of the unperturbed 
mean-field problem appearing in \eqref{H_c} is for a spherical hard-wall
potential simply given by the dipole matrix element 
$\langle\alpha|z|\beta\rangle$ \cite{weick05}.
Here, we have assumed that the surface plasmon oscillates in the $z$ direction.
The sum of \eqref{H_c} has to be understood with a low-energy cutoff since the 
electron-hole excitations of low energy do not couple to the surface plasmon 
\cite{weick06, yannouleas}. This cutoff energy has been estimated in
\cite{cesar} to be of the order of $3/5$ of the Fermi energy for the
case of Na nanoparticles. 

The decomposition \eqref{H_decomposition} is reminiscent of the well-stu\-died
case where the degree of freedom of interest (the surface plasmon in 
our case) is weakly coupled to a reservoir with many degrees of freedom
(the relative coordinates). Interestingly, it can be shown that it
is sufficient to have about $N=20$ conduction electrons to define a proper
environment for the surface plasmon\cite{cesar}. Since the electronic system 
is coupled to phonons, a time dependence of the electronic temperature $T$ may 
have to be taken into account. However, the physical parameters describing the 
dynamics of the surface plasmon are only weakly temperature dependent 
\cite{weick06}, and thus we can neglect the implicit time dependence of those 
parameters when $T$ is much smaller than the Fermi temperature of the system. 

If the nanoparticle is subject to a laser excitation, we have to add 
the coupling $H_{\rm F}(t)$ between the electrons and  the laser field 
to the Hamiltonian (\ref{H_decomposition}). Excitation frequencies 
$\omega_{\rm L}$ close to the plasmon resonance are located in the visible 
range, and therefore the electromagnetic field has a wavelength much 
larger than the size of the nanoparticle. This external electrical field 
can be considered as spatially homogeneous,
${\bf E}(t)=E_0\cos{(\omega_{\rm L}t)}{\bf e}_z$, and thus it only couples 
to the electronic centre of mass. In the dipolar approximation, the 
interaction between the driving field and the centre-of-mass system reads
\begin{equation}
\label{H_F}
H_{\rm F}(t) = 
(b^\dagger+b) \hbar \Omega_{\rm R} \cos{(\omega_{\rm  L}t)}\,. 
\end{equation}
The Rabi frequency $\Omega_{\rm R}$ is given by 
\begin{equation}
\label{Rabi_freq}
\hbar \Omega_{\rm R} = e E_0 N \ell\,,
\end{equation}
where 
\begin{equation}
\label{ell}
\ell=\sqrt{\frac{\hbar}{2N m_{\rm e}\tilde\omega_{\rm M}}}
\end{equation}
is the characteristic length associated with the harmonic oscillator \eqref{H_cm}.

Different physical regimes can be attained according to the relative 
values of the typical energies $\hbar \omega_{\rm M}$, $a\Lambda$,
$\hbar \Omega_{\rm R}$, and the Fermi energy of the metal.

%==============================================================================
%==============================================================================
%==============================================================================
%==============================================================================
\section{Centre-of-mass density matrix}
\label{sec_RDM}
In the first part of this section we consider the free evolution of the 
centre of mass without the external driving field. 
The driving term \eqref{H_F} will be added in the second part.

%==============================================================================
%==============================================================================
%==============================================================================
\subsection{Free evolution of the centre of mass}
\label{sec_free_evolution}
The evolution of the whole system in the absence of the electromagnetic field 
($H_{\rm F}=0$) can be expressed in terms of the total density matrix 
$W(t)$. Assuming that the coupling \eqref{H_c} is a small perturbation, we 
write the time evolution of the density matrix as
\begin{equation}
\label{equmotion}
\dot{ \tilde{W}}(t) = 
-\frac{{\rm i}}{\hbar} 
\left[ \tilde H_{\rm c}(t), \tilde W(t) \right]\,,
\end{equation}
where the tilde denotes operators in the interaction picture with respect to
the uncoupled Hamiltonian $H_{\rm cm}+H_{\rm rel}$.

As we are only interested in the dynamics of the centre of mass, we trace out
the relative coordinates to obtain the reduced density matrix $\rho = 
\text{Tr}_\text{rel} W$ which obeys the equation of motion
\begin{align}
\label{rho_cm}
\dot{ \tilde{\rho}}(t) =&
-\frac{{\rm i}}{\hbar} {\rm Tr}_{\rm rel} 
\left[ \tilde H_{\rm c}(t), W(0) \right]\nonumber\\
&-\frac{1}{\hbar^2} \int_{0}^t {\rm d}s \, {\rm Tr}_{\rm rel}  
\left[ \tilde H_{\rm c}(t), \left[ \tilde H_{\rm c}(s), 
\tilde W(s) \right] \right]\,. 
\end{align}

In dissipative quantum dynamics the environment is thought of as being infinite
and remaining in thermal equilibrium at all times, i.e., its intrinsic 
properties like the temperature do not change as a consequence of its coupling 
to the system  \cite{weiss}. In order to satisfy this condition, we have to 
assume that the internal equilibration of the relative degrees of freedom 
after an excitation occurs sufficiently fast. This last assumption is not 
equivalent to the thermalisation of the whole electron gas, which can only 
occur when the plasmon has decayed. The internal thermalisation hypothesis 
can be justified a posteriori by checking that the timescale of the 
temperature change of the relative degrees of freedom is larger than all the 
other relevant timescales. 

Restricting ourselves to the lowest order in the correlations between system 
and environment in \eqref{rho_cm}, the weak coupling limit that we assume to 
be valid allows to write (for details, see Chap.~4 in \cite{cohen})
\begin{equation}
\label{rho_0}
W(t)\approx\rho(t)\otimes\rho_{\rm rel}
\end{equation}
at any time $t\geqslant0$. Here, 
\begin{equation}
\label{rho_rel}
\rho_{\rm rel} = \frac{{\rm e}^{-\beta (H_{\rm rel}-\mu N)}}{\Xi} 
\end{equation}
represents the density matrix of the environment in the grand-canonical 
ensemble, i.e., we ignore for this purpose the finite nature of the electron 
gas. $\Xi$ is the grand-ca\-non\-i\-cal partition function at the inverse 
temperature $\beta=(k_{\rm B}T)^{-1}$ and $\mu$ is the chemical potential. 

As a consequence of the structure of the coupling $H_{\rm c}$, the fact that 
$d_{\alpha\alpha}=0$ due to the dipole selection rules \cite{weick05}, and the 
assumption of a factorising initial condition, the first term 
on the right-hand side of \eqref{rho_cm} vanishes. With the weak-coupling 
limit, \eqref{rho_cm} then reads
\begin{equation}
\label{markov}
\dot{\tilde{\rho}}(t)\simeq 
-\frac{1}{\hbar^2} \int_0^t {\rm d}s\, {\rm Tr}_{\rm rel} 
\left[ \tilde H_{\rm c}(t), \left[\tilde H_{\rm c}(s), 
\tilde \rho(s)\otimes\rho_{\rm rel} \right] \right]\,.
\end{equation}
The trace over the electronic environment \eqref{H_rel} yields
\begin{align}
\label{before_markov}
\dot{ \tilde{\rho}}(t) &=
\frac{1}{\hbar^2}\int_0^t{\rm d}\tau\; C(\tau)\nonumber\\
&\times\left[\tilde b^\dagger(t)+\tilde b(t), \tilde\rho(t-\tau)
\left(\tilde b^\dagger(t-\tau)+\tilde b(t-\tau)\right)\right]+{\rm h.c.}
\end{align}
The correlation function of the environment $C(\tau)$ is defined as
\begin{equation}
\label{correlation_function}
C(\tau)=\Lambda^2\sum_{\alpha\beta}\left[1-f(\varepsilon_\alpha)\right]
f(\varepsilon_\beta) \left|d_{\alpha\beta}\right|^2
{\rm e}^{{\rm i}\omega_{\alpha\beta}\tau}\,,
\end{equation}
and contains the relevant information on the time evolution of the
relative-coordinate degrees of freedom. In \eqref{correlation_function},
$f(\varepsilon)= [{\rm e}^{\beta (\varepsilon-\mu)}+1]^{-1}$ is the 
Fermi function and 
$\omega_{\alpha\beta}=(\varepsilon_\alpha-\varepsilon_\beta)/\hbar$.

For not too small nanoparticles, the electronic environment described by the 
Hamiltonian \eqref{H_rel} contains a large number of degrees of freedom, and 
its spectrum is therefore quasi-continuous. Thus, the exponentials 
contributing to the correlation function \eqref{correlation_function} very 
efficiently suppress the sum for not too small $\tau$. Indeed, for zero 
temperature it is shown in \cite{cesar} that the typical correlation time 
$\langle\tau_\text{cor}\rangle$ of the environment in terms of the surface 
plasmon relaxation time $\tau_{\rm pl}$ is given by 
$\langle\tau_\text{cor}\rangle/\tau_\text{pl}\approx (k_\text{F}a)^{-1}$, 
where $k_{\rm F}$ is the Fermi wavevector. For nanoparticles with a radius
larger than approximately \unit[1]{nm}, one typically finds $\tau_\text{pl}$ 
to be of the order of or larger than $10\langle\tau_\text{cor}\rangle$. The 
effect of finite temperatures on this estimate is expected to be weak 
\cite{weick06}.

For sufficiently large nanoparticles, we can therefore work within the
Markov approximation \cite{cohen}, where the density matrix $\rho$ in
\eqref{before_markov} can be assumed to vary on much longer timescales
than the decay time of the correlation function $C(t)$. The integral in
\eqref{before_markov} can then be carried out. After returning to the 
Schr{\"o}dinger picture, and within the secular approximation \cite{cohen}, 
where highly oscillating terms are neglected, we finally obtain the master 
equation for the reduced density matrix of the centre-of-mass degree of 
freedom in the absence of the external driving field
\begin{align}
\label{master_equation_nodriving}
\dot{\rho}(t)=&-{\rm i}\omega_{\rm sp}\left[b^\dagger b,
\rho(t)\right]\nonumber\\
&-\frac{\gamma_-}{2}\left[b^\dagger b\rho(t)+\rho(t) b^\dagger b
-2b\rho(t) b^\dagger\right]\nonumber\\
&-\frac{\gamma_+}{2}\left[b b^\dagger\rho(t)+\rho(t) b b^\dagger
-2b^\dagger\rho(t) b\right]\,.
\end{align}
Here, we have defined 
\begin{equation}
\label{gamma_pm}
\gamma_\pm=\frac{2\pi}{\hbar^2}\Lambda^2\sum_{\alpha\beta}
\left[1-f(\varepsilon_\alpha)\right]f(\varepsilon_\beta)
\left|d_{\alpha\beta}\right|^2
\delta\left(\tilde\omega_{\rm M}\pm\omega_{\alpha\beta}\right)\,,
\end{equation}
while the renormalised surface plasmon frequency reads
\begin{equation}
\label{omega_sp}
\omega_{\rm sp}=\tilde\omega_{\rm M}-\delta
\end{equation}
with
\begin{equation}
\label{delta}
\delta =\frac{2}{\hbar^2}\Lambda^2{\cal P}\sum_{\alpha\beta}
\left[1-f(\varepsilon_\alpha)\right]f(\varepsilon_\beta)
\left|d_{\alpha\beta}\right|^2
\frac{\omega_{\alpha\beta}}{\omega_{\alpha\beta}^2-\tilde\omega_{\rm M}^2}\,,
\end{equation}
where $\cal P$ denotes the Cauchy principal value.

The expressions for $\gamma_\pm$ and $\delta$ entering
\eqref{master_equation_nodriving} have simple physical 
interpretations\cite{weick06}: Using Fermi's golden rule, $\gamma_+$ and 
$\gamma_-$ are related to the lifetime 
$\tau_n=\gamma_n^{-1}$ of the $n^{\rm th}$ excited state 
of the harmonic oscillator describing the centre-of-mass system by 
\begin{equation}
\label{gamma_n}
\gamma_n=n\gamma_-+(n+1)\gamma_+\,, 
\end{equation}
while $\delta$ is the frequency shift due to the interaction of the system 
with the electronic environment. As the surface plasmon corresponds to the
first excited state of the harmonic oscillator, its lifetime is given
by $\tau_{\rm pl}=\gamma_1^{-1}$. We have evaluated these important and 
experimentally relevant quantities in references~\cite{weick05} and 
\cite{weick06} as a function of the size and temperature of the nanoparticle. 
The shift $\delta$ is positive, and thus we have a redshift of the surface 
plasmon frequency in addition to the redshift induced by the 
spill-out effect described by \eqref{omega_M}. This redshift of the surface 
plasmon frequency is due to the coupling of the centre-of-mass system to the
electronic environment. It is analogous to the 
Lamb shift known in atomic physics \cite{lamb, cohen}.

%==============================================================================
%==============================================================================
%==============================================================================
\subsection{Effect of the laser}
\label{sec_driving}
We now consider the role of an external driving field whose interaction with 
the centre-of-mass coordinate is described by the time-dependent Hamiltonian 
$H_{\rm F}(t)$ defined by \eqref{H_F}. In a first
step, we neglect the coupling Hamiltonian $H_{\rm c}$ between the 
centre-of-mass and the electronic environment during the excitation
process. Thus, we do not have any relaxation mechanism for the surface plasmon 
excitation, and its frequency \eqref{omega_sp} remains unrenormalised. 
As a consequence, we have
\begin{equation}
\label{master_equation_driving}
\dot\rho(t)=-{\rm i}\tilde\omega_{\rm M}\left[b^\dagger b, \rho(t)\right]
-{\rm i}\Omega_{\rm R}\cos{\left(\omega_{\rm L}t\right)}
\left[b^\dagger+b, \rho(t)\right]
\end{equation}
which describes the standard time evolution of a harmonic oscillator driven 
by an external mo\-no\-chro\-ma\-tic field. 

Assuming that the driving does not influence the dissipation of the
surface plasmon excitation, we can add the contributions from 
$H_{\rm c}$, i.e., the dissipative part of the reduced density matrix 
described by the master equation \eqref{master_equation_nodriving}, and 
from $H_{\rm F}(t)$ (see Eq.~\eqref{master_equation_driving}) 
independently. This is justified provided that the memory time 
of the environment $\langle\tau_{\rm cor}\rangle$ is much smaller than 
$\Omega_{\rm R}^{-1}$ which is determined by the coupling to the driving field \cite{cohen}.
The matrix representation of the master equation in the harmonic oscillator 
basis is then that of a driven damped harmonic oscillator,
\begin{align}
\label{Bloch}
\dot{\rho}_{nm}=&-{\rm i}{\omega}_{\rm sp}(n-m) \rho_{nm}\nonumber\\
&-\gamma\left(\frac{n+m}{2}\rho_{nm}-\sqrt{(n+1)(m+1)}\rho_{n+1, m+1}\right) 
\nonumber\\
&-{\rm i}\Omega_{\rm R}\cos{\left(\omega_{\rm L}t\right)} 
\left(\sqrt{n}\rho_{n-1,m}+\sqrt{n+1}\rho_{n+1,m}\right.\nonumber\\
&-\left.\sqrt{m}\rho_{n,m-1}-\sqrt{m+1}\rho_{n,m+1}\right)\,.
\end{align}
In this equation, we have neglected $\gamma_+$ as compared to $\gamma_-$
and set $\gamma=\gamma_-$ which corresponds to the Landau damping linewidth. 
Indeed, it is easy to show from \eqref{gamma_pm} 
that the rate $\gamma_+$ is related to $\gamma_-$ through the 
detailed-balance relation 
$\gamma_+={\rm e}^{-\beta\hbar\tilde\omega_{\rm M}}\gamma_-$.
We have $\gamma_+\ll\gamma_-$ for temperatures up to a few thousand degrees
since $\hbar\tilde\omega_{\rm M}$ is of the order of several eV.

%==============================================================================
%==============================================================================
%==============================================================================
%==============================================================================
\section{Two-level system approach}
\label{sec_2l}
In order to obtain from \eqref{Bloch} the time evolution of the centre-of-mass 
degree of freedom under the influence of the external driving field and the 
coupling to the relative-coordinate system, it is useful to introduce an 
appropriate simplification. The centre-of-mass system, which has been modelled
as a harmonic oscillator described by the Hamiltonian \eqref{H_cm}, can be 
truncated to a two-level system. The applicability of such an approximation is 
discussed in the appendix but can be motivated as follows: Except for a very 
strong driving field, the harmonic oscillator states 
above the first excited state are not significantly populated. Furthermore, 
the detuning between the frequency of the laser and the resonance frequency of 
the system plays in favour of the two-level description. Moreover, there exist 
additional damping mechanisms which are not included in our model which tend
to depopulate the higher states, like the ionisation of an electron via the 
double plasmon state \cite{weick05} or the radiation damping 
\cite{brack_RMP}. In the case of pump-probe experiments on noble-metal 
nanoparticles embedded in a dielectric medium, interactions with the 
surrounding matrix provide further decay channels, e.g. via coupling to 
phonons, localised states, surface states, etc. As the second collective 
level, the so-called double plasmon, has a width which is significantly 
larger than the one of the simple surface plasmon, it is justified to 
neglect all excited levels but the first one. In addition, the 
double-plasmon state has not been clearly identified 
experimentally, even though indirect observations of such a state have been 
reported in experiments on charged sodium clusters in vacuum\cite{schlipper}.

Writing the master equation \eqref{Bloch} for the two collective
states $|0\rangle$ and $|1\rangle$, introducing the new variables 
\begin{equation}
\label{hat_rho}
\hat{\rho}_{nm}=\rho_{nm}{\rm e}^{{\rm i}\omega_{\rm L}(n-m)t}
\end{equation}
and keeping only the terms which significantly contribute close to the 
resonance $\omega_{\rm L}\approx\omega_{\rm sp}$ (rotating wave approximation 
\cite{cohen}), one obtains 
\begin{subequations}
\label{RWA}
\begin{align}
\dot{\hat{\rho}}_{11}&={\rm i}\frac{\Omega_{\rm R}}{2}\left(\hat\rho_{10}-\hat\rho_{01}\right)
-\gamma\hat\rho_{11}\,,\\
\dot{\hat{\rho}}_{01}&=-{\rm i}\delta_{\rm L}\hat\rho_{01}
-{\rm i}\frac{\Omega_{\rm R}}{2}\left(\hat\rho_{11}-\hat\rho_{00}\right)
-\frac{\gamma}{2}\hat\rho_{01}\,. 
\end{align}
\end{subequations}
Here, $\delta_{\rm L}=\omega_{\rm L}-\omega_{\rm sp}$ is the detuning between 
the laser and the resonance frequency, and we have the conditions 
$\hat\rho_{00}+\hat\rho_{11}=1$ and $\hat\rho_{10}^\ast=\hat\rho_{01}$. 
It is often useful to define a scaled 
detuning $\Delta=\delta_{\rm L}/\gamma$ and the saturation parameter
\begin{equation}
\label{s}
s=2\left(\frac{\Omega_{\rm R}}{\gamma}\right)^2
\end{equation}
which is a measure of the ratio between the field intensity and the damping
strength. As a
function of these parameters, whose experimentally relevant values are 
discussed in the next section, the stationary solutions of the 
Bloch equations \eqref{RWA} are \cite{cohen}
\begin{align}
\label{rho_11_st}
\rho_{11}^{\rm st} &= \frac{s}{2(1+s+4\Delta^2)}\,, \\
\label{rho_01_st}
\rho_{01}^{\rm st}(t) &= {\rm e}^{{\rm i} \omega_{\rm L}t}
\frac{2 \Delta + {\rm i}}{1+s+4\Delta^2}\left(\frac{s}{2}\right)^{1/2}\,. 
\end{align}

The occupation probability $\rho_{11}^\text{st}$ of the first excited state
of the centre-of-mass system increases with increasing saturation parameter
$s$ and decreases with increasing detuning $\Delta$ 
between the frequency of the laser field $\omega_{\rm L}$ and the resonance 
frequency $\omega_{\rm sp}$. The coherence $\rho_{01}$ determines the mean 
centre-of-mass coordinate according to 
$\langle Z\rangle=2\ell\textrm{Re}(\rho_{01})$ where $\ell$ has been defined 
in \eqref{ell}. The amplitude of the oscillation of $\langle Z\rangle$ 
increases with increasing $s$ up to $s=1+4\Delta^2$, while it decreases for
larger $s$ and approaches zero in the limit of infinite $s$.

For zero detuning, $\delta_\text{L}=0$, \eqref{RWA} can be solved analytically.
Assuming that the system is initially in its ground state, 
$\rho(0) = \vert0\rangle\langle0\vert$, the full time dependence of the density
matrix is determined by \cite{walls}
\begin{align}
\label{rho_11}
\rho_{11}(t)=&\ \frac{s}{2(1+s)} 
\left\{ 
1-{\rm e}^{-3 \gamma t/4}
\left[
\cosh{\left( \frac{\gamma t}{4} \sqrt{1-8s} \right)}\right.\right.\nonumber\\
&\left.\left.+\frac{3}{\sqrt{1-8s}}\sinh{\left( \frac{\gamma t}{4} \sqrt{1-8s} \right)}
\right]
\right\}
\end{align}
for the population of the excited state, and 
\begin{align}
\label{rho_01}
\rho_{01}(t) =&\ \frac{{\rm i}{\rm e}^{{\rm i}\omega_{\rm sp}t}}
{1+s}\left(\frac{s}{2}\right)^{1/2}
\left\{
1-{\rm e}^{-3 \gamma t/4} 
\left[
\cosh{\left( \frac{\gamma t}{4} \sqrt{1-8s} \right)}\right.\right.\nonumber\\
&\left.\left.+\frac{1-2s}{\sqrt{1-8s}}\sinh{\left( \frac{\gamma t}{4} \sqrt{1-8s} \right)}
\right]
\right\}
\end{align}
for the coherence. 

\begin{figure}[t]
\begin{center}
\includegraphics[width=\columnwidth]{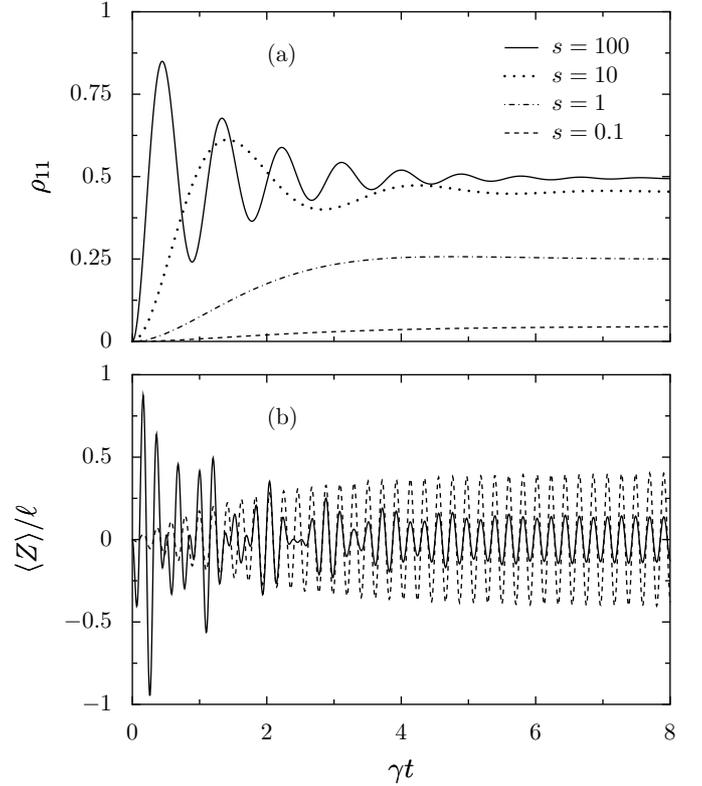}
\caption{\label{fig_rho_11}
(a) Population of the excited state $\rho_{11}$
\eqref{rho_11} as a function of $\gamma t$ for different values of the 
saturation parameter $s$ and for $\delta_{\rm L}=0$.
(b) Mean centre-of-mass coordinate $\langle Z\rangle$ for 
$s=0.1$ and 100. In the figure, $\omega_{\rm sp}/\gamma=30$ and 
$\delta_{\rm L}=0$.}
\end{center}
\end{figure}

Figure \ref{fig_rho_11}a shows the population \eqref{rho_11} of the excited 
state $\rho_{11}$ as a function of time for various values of the
saturation parameter and without detuning ($\delta_{\rm L}=0$). 
For $s\leqslant1/8$, $\rho_{11}$ monotonically increases 
as a function of time to reach the stationary value given in \eqref{rho_11_st}. 
For $s>1/8$, the centre-of-mass system exhibits damped Rabi 
oscillations and reaches the steady state on the timescale 
$\sim \gamma^{-1}$.
If the laser field is turned off after a certain time $\tau$, the
population of the excited state relaxes to zero according to
\begin{equation}
\rho_{11}(t)=\rho_{11}(\tau){\rm e}^{-\gamma (t-\tau)}\,,
\end{equation}
while the coherence of the system decays as 
\begin{equation}
\rho_{01}(t)=\rho_{01}(\tau)
{\rm e}^{-(\gamma/2-{\rm i}\omega_{\rm sp})(t-\tau)}\,.
\end{equation}
One can see in Figure \ref{fig_rho_11} that even for a strong saturation
parameter, the steady state regime is reached for times $t\lesssim8\gamma^{-1}$.

In Figure \ref{fig_rho_11}b, we show the mean centre-of-mass position  
$\langle Z \rangle$ as a function of time. For $s\ll1$, $\langle Z\rangle$ 
oscillates regularly, while its amplitude increases monotonically as a 
function of time. In the case of strong saturation parameter $s\gg 1$, it 
is clear from \eqref{rho_01} that the dynamics of the centre of mass is 
determined by the beating between the two oscillating contributions at the 
frequencies $\omega_{\rm sp}$ and $\Omega_{\rm R}$ for times shorter than a 
few $\gamma^{-1}$ (see the solid line in Fig.~\ref{fig_rho_11}b).

%==============================================================================
%==============================================================================
%==============================================================================
\section{Saturation parameter}
\label{sec_s}
The results in the previous section show that the saturation parameter 
\eqref{s} is decisive for the dynamics of the surface plasmon excitation in 
the presence of an external driving field. In typical photoabsorption 
experiments \cite{deheer}, a weak laser field excites an ensemble of 
nanoparticles. For a laser power of a few mW, the resulting electric field 
is of the order of \unit[$10^3$]{Vm$^{-1}$}, and thus the Rabi frequency 
\eqref{Rabi_freq} is no more than several $\mu$eV. The order of magnitude of 
the surface plasmon linewidth in metallic nanoparticles is \unit[100]{meV}, so that 
one finds $s\ll1$. In that case we are in the linear-response regime, i.e., 
the power absorbed by the nanoparticle is proportional to the intensity of 
the laser.

On the contrary, in pump-probe experiments, the na\-no\-par\-ti\-cles are 
excited by an ultrashort pump laser pulse of high intensity. As a typical 
example, we consider the experiments of \cite{bigot2} on silver nanoparticles 
with an average radius $a=\unit[3.25]{nm}$ embedded in a glass matrix. 
These nanoparticles present a broad absorption spectrum around the resonance frequency 
$\omega_{\rm sp}=\unit[2.85]{eV}$. According to time-dependent local
density approximation calculations \cite{weick05}, the width of this resonance 
is approximately $\gamma=\unit[50]{meV}$. The estimation of 
the Rabi frequency \eqref{Rabi_freq} requires the knowledge of the electrical 
field $E_0$ of the laser. This quantity can be inferred from the energy 
$\bar u=\epsilon_0\epsilon_{\rm r}E_0^2c\tau/2$ of a laser pulse divided by 
its cross section \cite{bigot2}, where $c$ is the speed of light, $\epsilon_0$ 
the vacuum permittivity, $\epsilon_{\rm r}\simeq 6$ for glass, and $\tau$ the duration of 
the pulse (typically \unit[100]{fs}). With $\bar u=\unit[1]{Jm^{-2}}$, one 
finds $E_0=3.6\times \unit[10^7]{Vm^{-1}}$. 
The Rabi frequency is therefore $\Omega_{\rm R}\simeq\unit[0.38]{eV}$, 
which results in a very large saturation parameter $s \simeq 100$. Thus we are 
far beyond the linear-response regime. Notice that the duration of the 
pump-laser pulse considered here corresponds to $\gamma\tau\simeq8$. Thus, 
according to Figure~\ref{fig_rho_11}, even for ultrashort laser pulses the 
centre of mass can reach its steady state before the end of the pulse.

%==============================================================================
%==============================================================================
%==============================================================================
%==============================================================================
\section{Absorption spectrum of a weak probe beam}
\label{sec_abs}
As it is shown in the appendix, the dynamics of the surface
plasmon is well described by a two-level system driven by an external
field and damped by its coupling to the electronic environment. 
This is reminiscent of the well-known situation in quantum 
optics, where a transition between two levels of an atom is excited by a 
strong laser. The concept of dressed levels \cite{cohen, reynaud}, arising as 
coherent superpositions of atom-photon states has proven to be very useful to 
describe the coupled system, in particular when the driving is strong as 
compared to the damping. In the presence of a small detuning 
$\delta_{\rm L}$ between 
the pump laser field and the transition frequency of the two-level system, 
the absorption spectrum of a second laser acting as a weak probe contains 
sidebands. Interestingly, in one of the sidebands the probe beam is damped by 
absorption processes, while stimulated emission processes amplify the 
probe beam in the other sideband. This phenomenon has been observed 
experimentally for different driven systems. Examples include the 
case of an optical transition in sodium atoms \cite{wu77} and a nuclear 
magnetic hyperfine transition of a nitrogen-vacancy 
centre in diamond \cite{wei94}.

\begin{figure}[t]
\begin{center}
\includegraphics[width=\columnwidth]{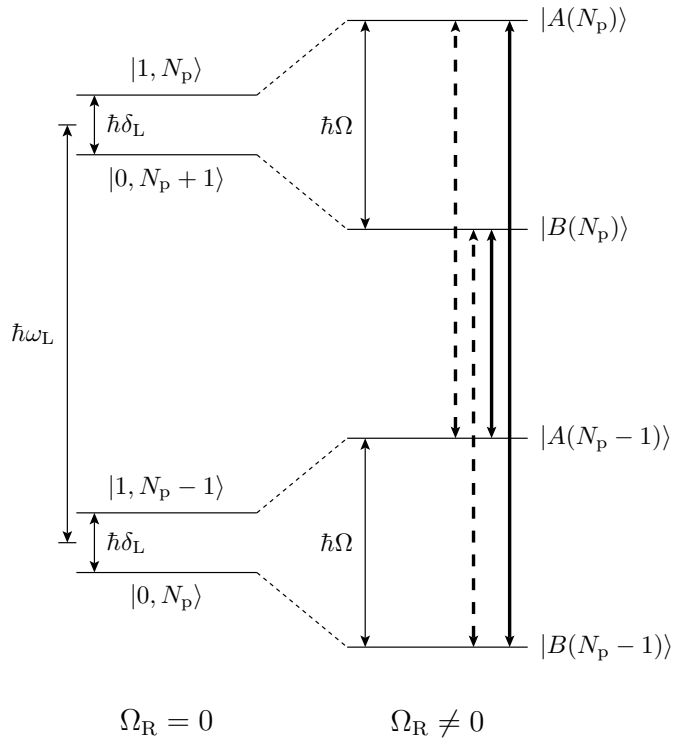}
\caption{\label{fig:dressed}
Scheme of the uncoupled ($\Omega_{\rm R}=0$) and dressed states 
($\Omega_{\rm R}\neq0$). The four allowed optical transitions are indicated by
vertical thick arrows. See the text for details.}
\end{center}
\end{figure}

In the present paper we focus on a surface plasmon excitation driven 
by a strong laser field. This is a completely analogous situation since 
the dynamics of the surface plasmon can be described by an effective 
two-level system with a transition energy $\hbar \omega_{\rm sp}$, 
coupled to a pump laser with frequency $\omega_\mathrm{L}$. 
Without coupling ($\Omega_\text{R}=0$), the basis states are 
$|0,N_\mathrm{p}+1\rangle$ and $|1,N_\mathrm{p}\rangle$ representing the 
plasmon in its ground and excited state with $N_\mathrm{p}+1$ and 
$N_\mathrm{p}$ photons in the pump laser field, respectively. 
The energy difference between those states is given by the detuning 
$\hbar\delta_\mathrm{L}=\hbar (\omega_\mathrm{L}-\omega_{\rm sp})$ of the pump 
laser with respect to the plasmon transition. The coupling by $H_{\rm F}$ 
(Eq.~\eqref{H_F}) leads to a mixing of the basis states, giving rise to the 
dressed plasmon levels with an energy splitting 
$\hbar\Omega=\hbar\sqrt{\Omega_\mathrm{R}^2+\delta_{\rm L}^2}$, where
$\Omega_\mathrm{R}$ is the Rabi frequency \eqref{Rabi_freq}. 
The dressed plasmon levels $|A(N_\mathrm{p})\rangle$ and 
$|B(N_\mathrm{p})\rangle$ arise as superpositions of the uncoupled basis 
states $|0,N_\mathrm{p}+1\rangle$ and $|1,N_\mathrm{p}\rangle$ referring to 
different photon numbers. In this doublet, we denote the upper and lower 
dressed level by $|A(N_{\rm p})\rangle$ and $|B(N_{\rm p})\rangle$, 
respectively. 

Four optical transitions (see thick arrows in Fig.~\ref{fig:dressed}) are 
allowed between dressed states corresponding to neighbouring photon numbers 
because of the mixture of those states with the ones of the uncoupled basis. 
Since the stationary populations of $|A(N_\mathrm{p})\rangle$ and 
$|B(N_\mathrm{p})\rangle$ are to a very good approximation independent of
$N_\text{p}$, there is no peak in the absorption spectrum of the weak probe 
beam at the frequency $\omega_{\rm L}$ (see the two thick dashed arrows in 
Fig.~\ref{fig:dressed}).

For $\delta_\mathrm{L}>0$, the upper dressed levels $|A(N_\mathrm{p})\rangle$ 
have a higher stationary population than the lower ones 
$|B(N_\mathrm{p})\rangle$. Therefore, a population inversion appears in the 
transition $|A(N_\mathrm{p})\rangle \leftrightarrow |B(N_\mathrm{p}-1)\rangle$.
This leads to a sideband at the frequency $\omega_\mathrm{L}+\Omega$ where the 
probe beam is amplified by stimulated emission. For the transition
$|B(N_\mathrm{p})\rangle \leftrightarrow |A(N_\mathrm{p}-1)\rangle$ the 
lower state has the higher population such that the sideband at frequency 
$\omega_\mathrm{L}-\Omega$ is of absorbing character. For negative detuning 
$\delta_\mathrm{L}<0$, when the pump laser frequency is below the bare 
transition energy, the populations of the dressed states are inverted. Now, 
the populations of the upper dressed levels 
$|A(N_\mathrm{p})\rangle$ are lower than the populations of the lower levels.
In this case, the sideband at the frequency $\omega_\mathrm{L}+\Omega$ is 
absorbing while the probe beam is expected to be amplified around the 
frequency $\omega_\mathrm{L}-\Omega$. 

The width of the sidebands is of the order of $\gamma$. In the 
so-called secular limit $\Omega \gg \gamma$, the sidebands 
are therefore well resolved and should be observable experimentally. 
Within this limit, it is straightforward to calculate the stationary 
populations of the dressed states \cite{reynaud}. For the case of a 
resonant pump laser $\delta_\mathrm{L}=0$, all the dressed states have 
the same populations. Therefore, to zero order in $\gamma/\Omega$, the 
transitions do not influence the probe beam. A more detailed discussion
beyond the parameter regimes considered here is given e.g. in 
references~\cite{wu77} and \cite{mollow72}.

An experimental observation of the sidebands should be feasible using an 
intense pump laser with $s\simeq 100$ and a pulse duration longer than 
\unit[100]{fs}. Then, the frequency width of the pump pulse is below 
\unit[0.01]{eV}, which allows to adjust the detuning 
$\delta_\mathrm{L}$ with high precision. With the Rabi frequency  
$\Omega_{\rm R}\simeq\unit[0.38]{eV}$ and the linewidth 
$\gamma=\unit[50]{meV}$ estimated in Section \ref{sec_s} for the
experiments of reference~\cite{bigot2}, one should be able to observe 
well resolved sidebands.

%==============================================================================
%==============================================================================
%==============================================================================
%==============================================================================
\section{Conclusion}
\label{sec_ccl}
Using standard methods of quantum optics, we have studied the dynamics of
the surface plasmon in metallic nano\-particles driven by an electrical 
field. 
Decoherence and dissipation occur due to the coupling of the collective
excitation to the electronic environment constituted by particle-hole pairs. 
Exploiting the Markovian character of the environment \cite{cesar}, we have 
established the master equation for the surface plasmon density matrix. 

We have shown that for realistic experimental situations, a model taking into
account only the two lowest levels is sufficient to describe the dynamics of 
the collective degree of freedom. This allows to consider a large variety of
experimentally relevant situations, from photoabsorption to pump-probe
setups.

Motivated by the analogy to quantum optics, we have introduced collective 
states that are dressed by the photons of the pump laser. We predict that
for strong driving, the dressed states will manifest themselves through
the appearance of sidebands in the absorption spectrum of a weak probe laser. 
Depending on the detuning of the pump laser, these spectral features 
have absorbing or amplifying character and their position relative to the
pump frequency is determined by the pump intensity. Although these
sidebands have so far not been observed in metallic nanoparticles, 
their detection should be feasible with modern experimental techniques, 
thereby offering access to dressed surface plasmon states.

%===========================================================================
%===========================================================================
%===========================================================================
%===========================================================================
\begin{acknowledgement}
We thank S.\ Kohler for useful discussions. 
We acknowledge financial support from DAAD and \'Egide through the Procope
program, from the BFHZ-CCUFB and from the European Union within the MCRTN 
program.
GW thanks the Deutsche Forschungsgemeinschaft for financial support during the
final phase of this work.
\end{acknowledgement}

\appendix
%==============================================================================
%==============================================================================
%==============================================================================
%==============================================================================
%==============================================================================
%==============================================================================
%==============================================================================
%==============================================================================
\section{Three-level system}
\label{chapter_3level}
In this appendix we discuss the validity of a two-level approach for the 
description of the centre-of-mass system, whose Hamiltonian is given in 
\eqref{H_cm}. For this purpose, we consider a three-level system
and study the conditions under which its description can 
be safely reduced to that of a two-level system. A phenomenological master
equation description of the plasmon in terms of three levels has been
discussed by Liau et al.\ \cite{dephasing}. In that case, however, 
three levels where required to describe two-photon processes in two-pulse
second-order interferometry.

According to \eqref{gamma_n},
the Landau damping linewidth for the second level of the harmonic 
oscillator at zero temperature is $2\gamma$, where $\gamma$ is the 
width of the first excited state \cite{molina,weick05}. In addition, two
more mechanisms contribute to the damping of the second level: (i) 
first order processes which lead to the decay into the first level (like the 
radiation damping \cite{deheer, brack_RMP} or the coupling to the surrounding 
matrix \cite{molina}); (ii) second-order processes which result in the direct 
decay into the ground state (like the ionisation \cite{weick05}). 
We denote the damping rates associated with these two additional channels 
by $\gamma_1$ and $\gamma_2$, respectively. 

Within the Lindblad theory \cite{lindblad}, these channels are accounted
for by adding 
\begin{equation}
\langle n|
\sum_{a=1, 2}
\frac{\gamma_a}{2}\left(2L_a\rho L_a^\dagger-L_a^\dagger L_a\rho-\rho L_a^\dagger L_a\right)
|m\rangle\,.
\end{equation}
to the right-hand side of the master equation \eqref{Bloch}.
We choose the Lindblad operators as $L_1=|1\rangle\langle 2|$ and
$L_2=|0\rangle\langle 2|$, which lead to transitions between the centre-of-mass
states $|2\rangle$ and $|1\rangle$, and $|2\rangle$ and $|0\rangle$,
respectively. The rotating-wave approximation then yields the following 
set of coupled differential equations
\begin{align}
\label{bloch_3l}
\dot{\hat{\rho}}_{00} =&
-{\rm i} \frac{\Omega_{\rm R}}{2} \left(\hat \rho_{10}-\hat \rho_{01}\right) 
+ \gamma \hat \rho_{11} + \gamma_2 \hat \rho_{22}\,, \nonumber\\
\dot{\hat{\rho}}_{22} =&
-{\rm i} \frac{\Omega_{\rm R}}{2} 
\sqrt{2}\left(\hat \rho_{12}-\hat \rho_{21}\right) 
-\left(2 \gamma + \sum_{a = 1, 2} \gamma_a \right) 
\hat \rho_{22}\,, \nonumber\\
\dot{\hat{\rho}}_{01} =&
-{\rm i} \delta_{\rm L} \hat \rho_{01} 
-{\rm i} \frac{\Omega_{\rm R}}{2} \left(\hat \rho_{11}-\hat \rho_{00}-\sqrt{2}
\hat \rho_{02}\right)\nonumber\\
&-\frac{\gamma}{2} \hat \rho_{01} + \sqrt{2} \gamma \hat \rho_{12}\,, \nonumber\\
\dot{\hat{\rho}}_{12} =&
-{\rm i} \delta_{\rm L} \hat \rho_{12} 
-{\rm i} \frac{\Omega_{\rm R}}{2} \left[\hat \rho_{02} +\sqrt{2} \left(\hat
\rho_{22} - \hat \rho_{11} \right) \right] \nonumber\\
&-\frac 12 \left( 3 \gamma + \sum_{a = 1, 2} \gamma_a \right) \hat \rho_{12}\,, \nonumber\\
\dot{\hat{\rho}}_{02} =&
-2 {\rm i} \delta_{\rm L} \hat \rho_{02} 
-{\rm i} \frac{\Omega_{\rm R}}{2} \left(\hat \rho_{12}-\sqrt{2} \hat
\rho_{01}\right) \nonumber\\
&-\frac 12 \left( 2 \gamma + \sum_{a = 1, 2} \gamma_a \right) \hat \rho_{02}\,. 
\end{align}

\begin{figure}[t]
\begin{center}
\includegraphics[width=\columnwidth]{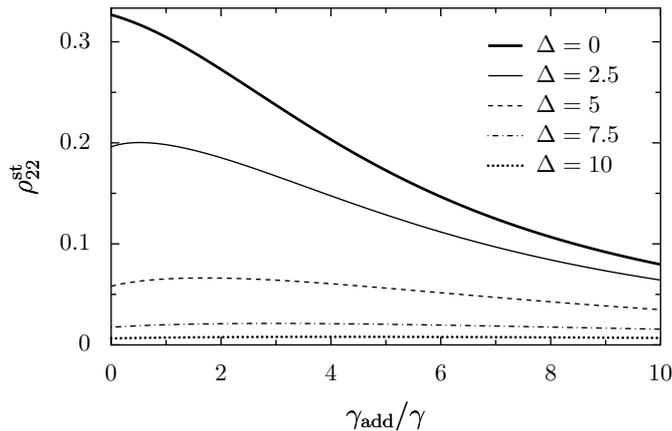}
\caption{\label{add_damping}Stationary 
solution of \eqref{bloch_3l} for the population of the second excited state as a
function of the additional damping constant $\gamma_{\rm add}$ and for various
values of the detuning $\Delta$. In the figure, the saturation parameter is
$s=100$.}
\end{center}
\end{figure}

In order to obtain quantitative results, the two additional damping constants 
in the following are assumed to be equal, i.e., $\gamma_1=\gamma_2=
\gamma_{\rm add}$. Figure \ref{add_damping} presents the stationary solution 
of \eqref{bloch_3l} for the population of the second excited state 
$\rho_{22}^{\rm st}$ as a function of $\gamma_{\rm add}$ for different values 
of the detuning $\Delta=\delta_{\rm L}/\gamma$ and a large saturation parameter
$s=100$. It can be seen that
for $\Delta\gtrsim5$, which corresponds to the typical detuning in the
pump-probe experiments of reference \cite{bigot2}, the probability to 
find the second collective state
occupied is less than 10\% and is almost constant as a function $\gamma_{\rm
add}$. In the case $\Delta<5$, $\rho_{22}^{\rm st}$ decreases with $\gamma_{\rm
add}$ but may be non-negligible. In this case, the assumption of a two-level 
system to describe the surface plasmon dynamics might be questionable.

For a rather small saturation parameter, i.e., for a weak external laser
field, the second collective state can be neglected for the description of 
the surface plasmon dynamics. In the
case of a large saturation parameter, as it is for example the case in
pump-probe experiments, this second collective state can be neglected for 
not too small detuning between the frequency of the pump laser and the 
resonance frequency.

%===========================================================================
%===========================================================================
%===========================================================================
%===========================================================================


\begin{thebibliography}{}

\bibitem{deheer}
W.A.\ de Heer, 
Rev.\ Mod.\ Phys.\ {\bf 65}, 611 (1993)

\bibitem{brack_RMP}
M.\ Brack, 
Rev.\ Mod.\ Phys.\ {\bf 65}, 677 (1993)

\bibitem{tokiz94}
T.\ Tokizaki, A.\ Nakamura, S.\ Kaneko, K.\ Uchida, S.\ Omi, H.\ Tanji, and Y.\
Asahara, 
Appl. Phys. Lett. {\bf 65}, 941 (1994)

\bibitem{bigot1}
J.-Y.\ Bigot, J.-C.\ Merle, O.\ Cregut, and A.\ Daunois,
Phys.\ Rev.\ Lett.\ {\bf 75}, 4702 (1995)

\bibitem{perner}
M.\ Perner, P.\ Bost, U.\ Lemmer, G.\ von Plessen, J.\ Feldmann, U.\ Becker, 
M.\ Mennig, M.\ Schmitt, and H.\ Schmidt, 
Phys.\ Rev.\ Lett.\ {\bf 78}, 2192 (1997)

\bibitem{halte99}
V.\ Halt\'e , J.\ Guille, J.-C.\ Merle, I.\ Perakis, and J.-Y. Bigot,
Phys.\ Rev.\ B {\bf 60}, 11738 (1999)

\bibitem{bigot2}
J.-Y.\ Bigot, V.\ Halt\'e, J.-C.\ Merle, and A.\ Daunois,
Chem.\ Phys.\ {\bf 251}, 181 (2000)

\bibitem{delfatti}
N.\ Del Fatti, F.\ Vall\'ee, C.\ Flytzanis, Y.\ Hamanaka, and A.\ Nakamura,
Chem.\ Phys.\ {\bf 251}, 215 (2000)

\bibitem{voisin}
C.\ Voisin, N.\ Del Fatti, D.\ Christofilos, and F.\ Vall\'ee, 
J.\ Phys.\ Chem. B {\bf 105}, 2264 (2001)

\bibitem{lett}
G.\ Weick, D.\ Weinmann, G.-L.\ Ingold, and R.A.\ Jalabert, Europhys.\ Lett.\
{\bf 78}, 27002 (2007)

\bibitem{gerchikov}
L.G.\ Gerchikov, C.\ Guet, and A.N.\ Ipatov,
Phys.\ Rev.\ A {\bf 66}, 053202 (2002)

\bibitem{weick05}
G.\ Weick, R.A.\ Molina, D.\ Weinmann, and R.A.\ Jalabert, 
Phys.\ Rev.\ B {\bf 72}, 115410 (2005)

\bibitem{weick06}
G.\ Weick, G.-L.\ Ingold, R.A.\ Jalabert, and D.\ Weinmann, 
Phys.\ Rev.\ B \textbf{74}, 165421 (2006)

\bibitem{cohen}
C.\ Cohen-Tannoudji, J.\ Dupont-Roc, and G.\ Grynberg, 
{\it Atom-photon interactions: Basic processes and applications} 
(Wiley-VCH, New York, 1992)

\bibitem{hagino}
K.\ Hagino, 
Phys.\ Rev.\ B {\bf 60}, R2197 (1999)

\bibitem{yannouleas}
C.\ Yannouleas and R.A.\ Broglia, 
Ann.\ Phys.\ (N.Y.) \textbf{217}, 105 (1992)

\bibitem{cesar}
C.\ Seoanez, G.\ Weick, R.A.\ Jalabert, and D.\ Weinmann, 
arXiv:cond-mat/0703720

\bibitem{weiss}
U.\ Weiss, 
{\it Quantum Dissipative Systems} 
(World Scientific, Singapore, 1993)

\bibitem{lamb}
W.E.\ Lamb and R.C.\ Retherford, 
Phys.\ Rev.\ {\bf 72}, 241 (1947);
W.E.\ Lamb, 
Rep.\ Prog.\ Phys.\ {\bf 14}, 19 (1951)

\bibitem{schlipper}
R.\ Schlipper, R.\ Kusche, B.\ von Issendorff, and H.\ Haberland, 
Phys.\ Rev.\ Lett.\ {\bf 80}, 1194 (1998);
Appl.\ Phys.\ A: Mater.\ Sci.\ Process.\ {\bf 72}, 255 (2001)

\bibitem{walls}
D.F.\ Walls and G.J.\ Milburn, 
\textit{Quantum Optics} 
(Springer-Verlag, Berlin, 1994)

\bibitem{reynaud}
C.\ Cohen-Tannoudji and S.\ Reynaud, 
J.\ Phys.\ B: Atom.\ Molec.\ Phys.\ {\bf 10}, 345 (1977);
C.\ Cohen-Tannoudji and S.\ Reynaud, in:
Proc.\ Int.\ Conf.\ on Multiphoton Processes, ed.\ by 
J.H.\ Eberly and P.\ Lambropoulos (Wiley, New York, 1978)

\bibitem{wu77}
F.Y.\ Wu, S.\ Ezekiel, M.\ Ducloy, and B.R.\ Mollow,
Phys.\ Rev.\ Lett.\ \textbf{38}, 1077 (1977)

\bibitem{wei94}
C.\ Wei and N.B.\ Manson,
Phys.\ Rev.\ A \textbf{49}, 4751 (1994)

\bibitem{mollow72}
B.R.\ Mollow, Phys.\ Rev.\ A \textbf{5}, 2217 (1972)

\bibitem{dephasing}
Y.-H.\ Liau, A.N.\ Unterreiner, Q.\ Chang, and N.F.\ Scherer,
J.\ Phys.\ Chem.\ B, {\bf 105}, 2135 (2001)

\bibitem{molina}
R.A.\ Molina, D.\ Weinmann, and R.A.\ Jalabert,
Phys.\ Rev.\ B {\bf 65}, 155427 (2002);
Eur.\ Phys.\ J.\ D {\bf 24}, 127 (2003)

\bibitem{lindblad}
G.\ Lindblad, 
Commun.\ Math.\ Phys.\ {\bf 48}, 119 (1976);
V.\ Gorini, A.\ Kossakowski, and E.C.G.\ Sudarshan, 
J.\ Math.\ Phys.\ {\bf 17}, 821 (1976) 

\end{thebibliography}
\end{document}